\documentstyle[epsfig]{l-aa}
\begin{document}
\def\fu{$f_1$}
\def\t{$\pm$}
\def\fd{$f_2$}
\def\fdu{$\phi_{21}$}
\def\fp{$f_1 + f_2$}
\def\fm{$f_2 - f_1$}
\def\cd{cd$^{-1}$}
\def\cds{cd$^{-1}$\,}
\def\kms{km~s$^{-1}$}
\def\kmss{km~s$^{-1}$\,}
\def\I{\'\i}
\def\salp{\vskip 0.3truecm}
\thesaurus{6(03.13.2, 08.12.3, 08.15.1, 08.22.1, 11.04.2)}
\title{The light curves of the short--period variable stars 
in the Carina dwarf Spheroidal galaxy}
\author{E. Poretti\inst{1}}
\institute {Osservatorio Astronomico di Brera, Via Bianchi 46,
I-23807 Merate, Italy\\E--mail: poretti@merate.mi.astro.it}
\offprints{E. Poretti}
\date{Received date; Accepted Date}
\maketitle
\markboth{E. Poretti: Short--period stars in the Carina galaxy}{ }
\begin{abstract}
The light curves of the high--amplitude, short--period variable stars
discovered by Mateo et al. (1998) in the Carina dwarf Spheroidal Galaxy
were re-analyzed to refine their properties and compare them with those
of galactic stars. In spite of a different pulsation mode established by
means of different $PLM$ relationships, the Fourier parameters did not
reveal a separation between the two groups; however, they are slightly
different from those of galactic stars. The possibility that some of  these
stars are double--mode pulsators is  suggested.
\end{abstract}

\section{Introduction}
In the recent years there has been a dramatic increase of the number of known
variable stars (galactic and extragalactic) thanks to the results obtained in
the framework of larg--scale projects as MACHO, EROS and OGLE.
As a complementary work, some teams started a survey of variable stars in
other Local Group Galaxies.
These projects mainly provide light curves and standard $B$ and $V$ magnitudes.
It is quite evident that
these curves have to be investigated by means of alternative analysis
in order to extract all the scientific information
they contain.

 Recently,
Mateo et al. (1998) reported on the discovery of 20 pulsating stars in the
Carina dwarf Spheroidal Galaxy. Their periods are ranging from 0.048 d to 
0.077 d and their amplitude is above 0.30 mag in $V$. Similar stars are known
also in the Galaxy and they are called  $\delta$ Sct  stars
if Pop. I objects or SX Phe stars if Pop. II objects; the old definition ``Dwarf
Cepheids" (without distinction between Populations) was dropped. The analysis
of their light curves by means of the Fourier decomposition
was firstly made by Antonello et al. (1986) and then continued by
Poretti et al. (1990).

After obtaining the Fourier parameters, we shall compare the light curves of the
pulsating variables in the Carina dSph galaxy with those of the galactic
variables.

\section{Period verification and refinement}
The mean magnitude of the variables was about $V$=23 and
even if observations were carried out with a 4--m telescope, the single
measurements are affected by an error of 0.06 mag or more. Since  the number of
points is 30 in the best cases (stars 1 to 12), the mean light curves are of
limited accuracy and large uncertainties remain.

To determine their period, Mateo et al.
(1998) used a method based on the minimum string (Dworetsky 1983), which may
not be optimum for noisy data sets with a limited amount of points.
\begin{figure}
\centerline{\psfig{file=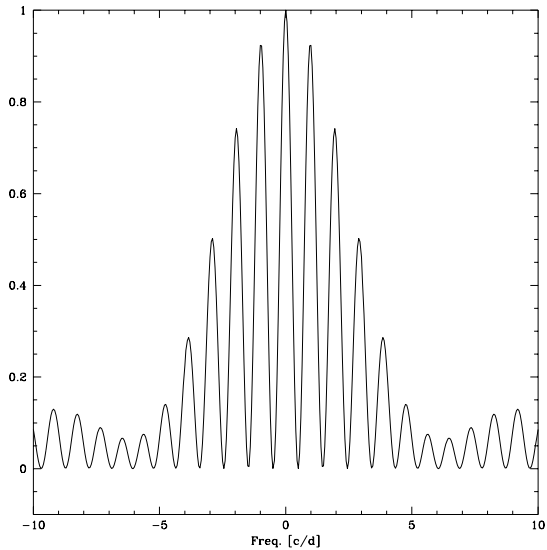,height=8truecm}}
\caption[]{The spectral window related to the data of the best observed
stars}
\end{figure}

Figure 1 shows the spectral window as obtained from the data related to stars
1--12, which were observed for a longer time in the two consecutive nights of
the sample. However, a night length of about 0.15 d is not sufficient to reduce
the aliases at $\pm$ 1 \cd and indeed their height is very similar to that of
the central peak; this fact clearly suggests that an ambiguity between the
selected  frequency and these aliases is always possible. 

To test this possible ambiguity, we repeated the analysis by using
the least--squares iterative
method (Vani\^cek 1971) to verify if changing the algorithm also changes the
results; Tab.~1 summarizes the results. In 9 cases the two approaches
detected the same frequencies, in 6 cases we preferred the $f+1$ \cd alias of
the term reported by Mateo et al. (1998), in 4
cases the $f-1$ \cd one; in one case (V~15) our method detected a peak related
to the $f-$3 \cd alias. As can be
seen from the columns listing the standard deviations, the fit is obviously
better in our approach since it is the goal of our search.

Some stars show a standard deviation greater than 0.10 mag (V~7,
V~8, V~17, V~20, V~15 and V~16), while the others cluster around 0.07 mag.
Such high values can be due to observational errors, but the possibility
that they (or part of them) could be  double--mode pulsators should be
investigated. Since the $F/1O$ double--mode 
pulsators are expected to be more easily observed, the fact that all the stars
(except the doubtful case of  V~16) with large residuals are in the group of
the $F$--mode pulsators is  a good hint.   
Unfortunately, owing to the limited data the frequency analyses could not give
convincing evidence of a second period.

Of course, it is not possible to state that the
frequencies listed in the right panel of Tab.~1 are the true ones, since also
in this case the
selection between adjacent peaks is very delicate and the scatter of
the light curves plays an important role; however, we can see what happens
by using these new refined values and looking at the corresponding  
$PLM$ relationships. Figure 2 shows how they changed. The solid lines for
fundamental ($F$; lower line) and first overtone ($1O$; higher line) pulsators
are calculated as in Fig. 5 in Mateo et al. (1998), i.e. by using m-M=20.09,
E($B-V$)=0.025 mag, [Fe/H]=--2.0; the dashed lines  represent the uncertainty
of 0.06 mag in the distance modulus.

As regard the $1O$ pulsators, the set of new frequencies gives related points
which are closer to the $PLM$ line; in particular Mateo et al. (1998) reported
an uncertain classification for the star V~2, which now seems to belong to
the group of 1$O$ pulsators. On the other
hand, the brightest star (V~14) still lies far from the line.
It is important to note that the five $1O$--pulsators are confirmed despite
the fact that a wrong choice between a peak and its $\pm$1\cd alias can 
introduce a horizontal shift of 0.03 (for $f$=15 \cd) in the $\log f$ value.

The components of the $F$--pulsator group are confirmed; note however that
in the Mateo et al.'s sample we have three stars with $1.1<\log f<1.12$,
while in our sample  only the star V~9 maintains such a long period. Hence,
the observed periods of the Carina galaxy stars are in average 
much shorter. Between the two groups the ``half--the--way" position of the
star V~16 remains unchanged; as noted above, it may be a double--mode
pulsator and its period only roughly known.

\begin{table*}
\begin{flushleft}
\caption{New frequency values for the variables in Carina dSph galaxy; if
no value is reported in the right part of the table, no improvement of the
fit was introduced by the new approach. Frequencies are in \cd; the standard
deviations of the fits are in 10$^{-2}$ mag; phase differences $\phi_{21}$
are in radians; $R_{21}$ is the ratio between the amplitudes of 2$f$ and $f$}
\begin{tabular}{rc |  rrrrr c rrrrr}
\hline
\noalign{\smallskip}
 &\multicolumn{1}{c}{ } & \multicolumn{5}{c}{Refined elements} & &\multicolumn{5}{c}{Mateo et al. (1998)}\\
\cline{3-7}   \cline{9-13}
\noalign{\smallskip}
 \multicolumn{1}{c}{Var.} & \multicolumn{1}{c}{$<V>$}& \multicolumn{1}{c}{Freq.}& \multicolumn{1}{c}{log~$f$}
& \multicolumn{1}{c}{s.d.} &  \multicolumn{1}{c}{$\phi_{21}$}&\multicolumn{1}{c}{$R_{21}$}
& &\multicolumn{1}{c}{Freq.}& \multicolumn{1}{c}{log~$f$}
& \multicolumn{1}{c}{s.d.} &  \multicolumn{1}{c}{$\phi_{21}$}&\multicolumn{1}{c}{$R_{21}$}\\
\noalign{\smallskip}
\hline
\noalign{\smallskip}
7 & 23.20 & 20.52 & 1.312 &11.3 & -- & -- \\
8 & 23.23 & 17.90 & 1.253 &11.4 & 3.20 & 0.24 &  & 18.94 & 1.277 &12.0 & 2.85 & 0.21 \\
4 & 23.02 & 17.60 & 1.246 & 8.8 & 4.49 & 0.53 &  & 16.69 & 1.222 &10.0 & 3.54 & 0.25 \\
10& 23.03 & 16.95 & 1.229 & 8.2 & 4.13 & 0.53 \\
5 & 23.06 & 16.92 & 1.228 & 7.2 & 4.08 & 0.14 \\
17& 22.95 & 15.40 & 1.187 &12.8 & 4.63 & 0.40 & & 16.58 & 1.219 &13.8 & 4.19 & 0.41  \\
13& 22.93 & 14.89 & 1.173 & 7.1 & 3.01 & 0.36 \\
20& 22.84 & 14.84 & 1.171 &10.6 & 4.19 & 0.39 \\
15& 22.97 & 14.45 & 1.160 &11.9 & 3.97 & 0.37 & & 17.30 & 1.238 &12.9 & 3.76 & 0.47 \\
6 & 22.79 & 14.29 & 1.156 & 7.1 & 4.07 & 0.43 \\
11& 22.75 & 13.95 & 1.145 & 7.3 & 3.58 & 0.39 & & 13.04 & 1.115 & 9.1 & 3.64 & 0.33 \\
3 & 22.92 & 13.90 & 1.143 & 7.1 & 5.06 & 0.43 & & 12.99 & 1.114 & 7.4 & 5.34 & 0.16  \\
9 & 22.79 & 13.26 & 1.122 & 5.6 & -- & -- \\
\noalign{\smallskip}
16& 22.81 & 15.38 & 1.187 &13.2 & 2.92 & 0.28 & & \\
\noalign{\smallskip}
2 & 23.10 & 21.90 & 1.340 & 7.7 & 4.57 & 0.33 & & 20.88 & 1.320 & 8.2 & 4.49 & 0.34\\
12& 22.83 & 18.05 & 1.256 & 5.4 & 4.39 & 0.28 & & 17.09 & 1.233 & 6.1 & 3.97 & 0.31  \\
18& 22.74 & 17.05 & 1.232 & 8.4 & 3.88 & 0.57 & & 18.12 & 1.258 & 8.8 & 3.82 & 0.57\\
14& 22.45 & 16.90 & 1.228 & 7.4 & 4.23 & 0.32 & & 15.97 & 1.203 & 8.4 & 3.99 & 0.34  \\
1 & 22.70 & 15.78 & 1.197 & 5.9 &  --   & --  & & 16.78 & 1.225 & 6.1 & --   & --    \\
19& 22.63 & 15.45 & 1.189 & 3.9 & 4.69 & 0.19 \\
\noalign{\smallskip}
\hline
\end{tabular}
\end{flushleft}
\end{table*} 
\begin{figure}
\centerline{\psfig{file=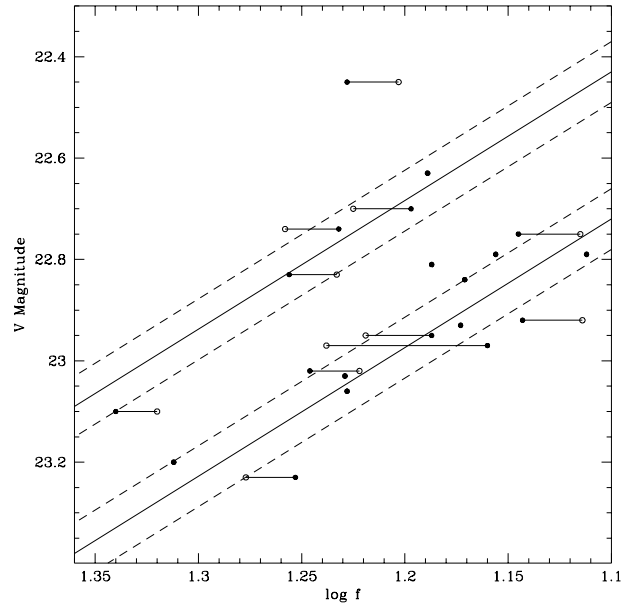,height=9truecm}}
\caption[]{The $PLM$ relationships for $F$--pulsators (lower line)
and $1O$--pulsators (upper line) are shown together with the frequency
values determined by Mateo et al. (1998; open circles) and
in the present work (filled circles). Horizontal lines connect points
related to the same star.}
\end{figure}

\section{Fourier parameters}
As a further step, we fitted the $V$ magnitudes  by means of the formula
\begin{equation}
V(t)= V_o + \sum_i {A_i \cos [2\pi~i~f  (t-T_o) +\phi_i ]}
\end{equation}
where $f$ is the frequency. From the least--squares coefficients we calculated
the Fourier parameters $R_{21}=A_{2f}/A_{f}$ and 
$\phi_{21}=\phi_{2f} - 2~\phi_{f}$. 
Typical error bars are $\pm$0.6 rad for \fdu values and $\pm$0.15 for the
$R_{21}$ ones. In three cases (V~1, V~7 and V~9) the
scatter and/or the phase coverage did not allow us to obtain a satisfactory
fit; hence, we cannot propose a reliable Fourier decomposition.

The goal was to obtain some hints about
the trends of such parameters when plotted versus the period. In the case of
Cepheids and Double--Mode Cepheids the $\phi_{21}$ parameter provide a
powerful tool to discriminate between fundamental ($F$) and first overtone
($1O$) pulsation modes. Recently, Musazzi et al. (1998) summarized the results
on the galactic stars, confirming the bimodal distribution of the $R_{21}$ 
ratio (although the need to increase the sample was emphasized) and the fast
increase of the $\phi_{21}$ values for periods between 0.05 and 0.10 d, followed
by a flattening for longer periods. 
All the short period variable stars discovered in the Carina galaxy lie in
the region where the fast increase of the \fdu values was observed.

In the case of the Carina galaxy  the approach to the discussion of 
light curves is the opposite as compared with 
our Galaxy. Indeed, the $P-L$ relationships allow us to separate in a
straightforward way the $F$--pulsators from the $1O$ ones, which is the final
goal of investigations in our Galaxy.

As regards the \fdu values and $1O$ pulsators, 
the range covered by the 5 stars is 3.88--4.69 rad; frequencies
are in the range 15.45--21.90 \cd. In the same frequency interval, the
$F$--stars display \fdu values ranging from 4.08 to 4.49 rad; only considering
the strong deviating point related to star V~8 this interval is extended to 3.20
rad as an upper limit.
As shown in Fig.~3, there is no clear separation between the two groups of
stars, at least toward short periods. More precise measurements
are necessary to find such a separation. This cannot be immediately interpreted
as a failure of the $\phi_{21}$ differences as mode discriminator, since 
in the case of Cepheids there is also a region where the  \fdu values
are very close each other even if the stars are pulsating in a different
mode. Perhaps 
a different behaviour could be seen toward longer period, but no such star
was observed up to now in the Carina galaxy.

We have four  galactic variable stars  in the same period range: CY Aqr, ZZ Mic, DY Peg
and V831 Tau. The Fourier decomposition of their light curves yields a very
narrow $\phi_{21}$ range between 3.66 and 3.81 rad, i.e. values smaller than
those observed in the Carina galaxy (Fig.~3). Explaining this difference is
far from obvious; for example, it cannot be related to the metallicity, since in
the Galaxy there is no
separation between Pop.~I and Pop.~II stars in the $\phi_{21}-P$ plot.
\begin{figure}
\centerline{\psfig{file=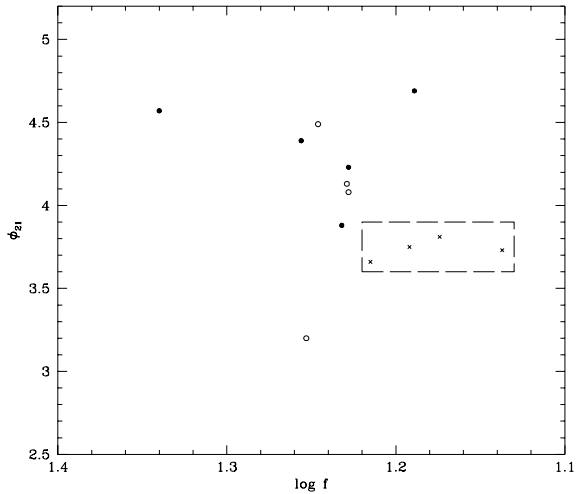,height=8.5truecm}}
\caption[]{\fdu values versus period for galactic pulsators (crosses),  Carina
galaxy 1$O$--pulsators (filled circles), Carina galaxy $F$--pulsators (open
circles). Note the narrow box of the galactic pulsators.}
\end{figure}

\section{Conclusions}
The re--analysis of the data reported by Mateo et al. (1998) allowed us to
obtain a partially different set of frequencies for the 20 short period
variable stars discovered in the Carina dSph galaxy. They are subdivided
into 6 $1O$--pulsators, 13 $F$--pulsators; 1 case remains uncertain.

The analysis
of the light curves was conducted star by star and the first Fourier parameters
were obtained for 17 variables. No clear separation between pulsation modes
is discernible in the \fdu$-P$ plot, but the very narrow range of the observed
periods hampered us to consider that statement as a definitive one. A more
extensive observational study is recommended, since a preliminary  comparison
with galactic variable stars displays small differences in the \fdu values.
Owing to the high standard deviation of the fit,
the possibility that some stars are double--mode pulsators is also suggested.

It should be also noted that a more accurate standard $V$ magnitude 
calibration is necessary. Indeed, the $V$ magnitudes used in the $PLM$
relationships were obtained by means of deep survey frames (longer
exposures, better S/N) and they can differ up to 0.15 mag from the values
obtained in the frames used to study the variability.

More accurate measurements  of these stars in the framework of a dedicated
project could greatly contribute to solve the matter, allowing us a more
reliable analysis. It is of paramount importance to quantify the differences
respect with the galactic stars since we could possibly relate them to
differences in physical quantities.
\begin{acknowledgements} The author wish to thank M.~Mateo for putting at his
disposal the data. J. Vialle improved the English form of the manuscript. 
\end{acknowledgements}

\end{document}